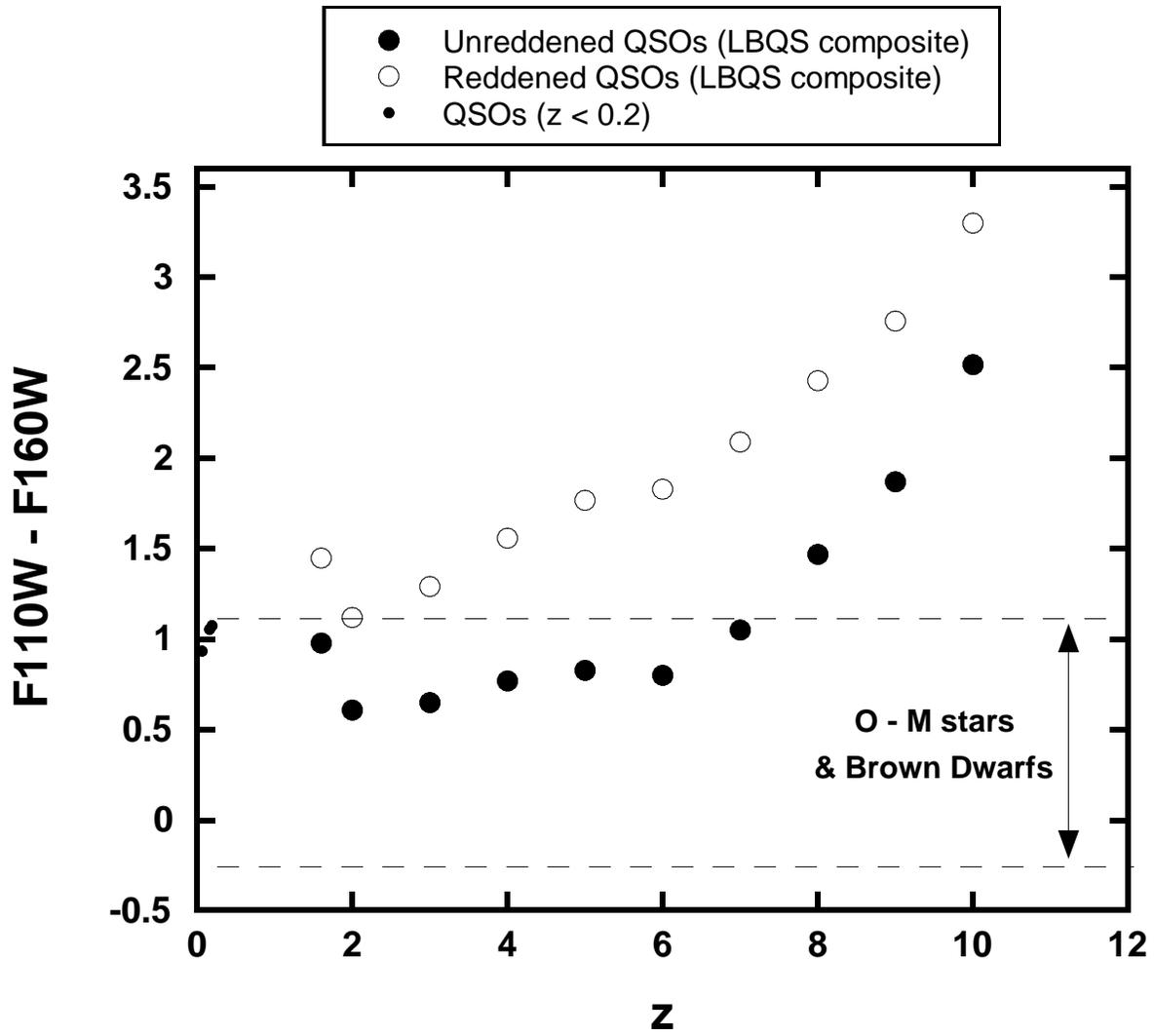

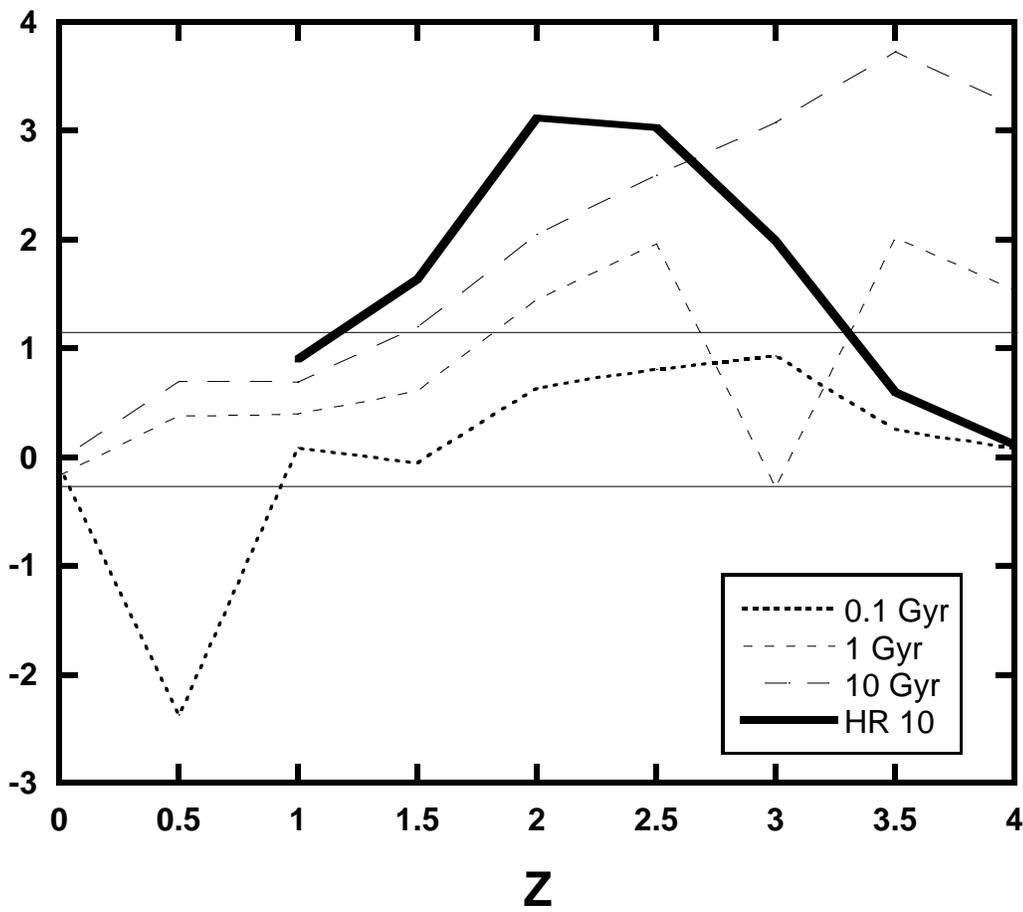

# A Color Analysis of the NICMOS Parallel Image Archive

Michael R. Corbin, Earl O'Neil,

Rodger I. Thompson, Marcia J. Rieke & Glenn Schneider

NICMOS Group, Steward Observatory, The University of Arizona

Tucson, AZ 85721; contact: mcorbin@as.arizona.edu

ABSTRACT

We present a photometric analysis of all high Galactic latitude ($|b| > 20°$) broad-band parallel images taken by the Near Infrared Camera and Multi-Object Spectrometer (NICMOS) instrument of the *Hubble Space Telescope* during its initial lifetime in HST Cycle 7. These images, taken through the F110W and F160W filters, reach a mean $3\sigma$ limiting magnitude of approximately 22 in both bands, and cover a total area of approximately 92 square arcminutes. The reddest of the 358 galaxies detected have F110W - F160W colors and F160W magnitudes consistent with being a combination of both dusty star-forming and evolved early-type galaxies at $1 < z < 2$. The surface density of these galaxies is comparable to that of the population of Extremely Red Objects (EROs) discovered in ground-based surveys ($\sim 100$ deg$^{-2}$), suggesting that EROs also represent a combination of both galaxy types in this redshift range. Roughly 10% of the detected galaxies appear to be blue compact dwarf galaxies at $z < 1$, a result consistent with studies of the HST Medium Deep Survey fields. The surface density of these objects down to a magnitude of 22 in F160W is $\sim 300$ deg$^{-2}$. None of the 631 point sources detected have F110W - F160W colors matching those expected for QSOs whose continua have been significantly reddened by internal dust. Our data limit the surface density of such QSOs to be $< 50$ deg$^{-2}$ down to the mean limiting magnitudes of the sample images. Since the surface density of QSOs detected on the basis of ultraviolet excess in optical surveys to comparable depth is $\sim 100$ deg$^{-2}$, this argues against the suggestion that dust-reddened QSOs comprise the undetected majority of the QSO population. The F110W - F160W color can also be used to identify unreddened QSOs at $z > 8$, but we find no such candidates. This is consistent with the evidence that QSO space density declines sharply at $z > 5$.

1. INTRODUCTION

An important benefit of the *Hubble Space Telescope* (HST) following its refurbishment in February 1997 has been the acquisition of several thousand images of random fields by the Near Infrared Camera and Multi-Object Spectrometer (NICMOS) and Space Telescope Imaging Spectrometer (STIS) instruments. These images, obtained in parallel with pointed observations by the other HST instruments, collectively offer "mini-surveys" of the sky at both medium depth and high spatial resolution at the wavelengths covered by NICMOS and STIS. The NICMOS parallel images in particular offer an unprecedented opportunity to investigate high-redshift galaxies and QSOs in the near infrared, as well as to probe the stellar populations of the Galaxy and other members of the Local Group.

In a companion paper (Corbin et al. 2000; hereafter Paper I) we identify and study the properties of 111 galaxies discovered in the NICMOS Camera 1 and Camera 2 parallel fields. Paper I joins other studies of galaxies in the NICMOS parallel images including Teplitz et al. (1998), Yan et al. (1998), McCarthy et al. (1999) and Treu & Stiavelli (1999). However, with the exception of Pasquali, DeMarchi & Freudling (1998), the properties of the stars and other point sources in these images have not been investigated. Additionally, while most of the broad-band NICMOS parallel images were obtained in Camera 3 and have corresponding grism images containing low-resolution spectra (see McCarthy et al. 1999), it is often impossible to extract these spectra for the faintest objects in the images, which are likely to be the most astronomically interesting. However, as will be discussed in the following sections, even the single color index that the broad-band parallel images provide can reveal such astrophysically important sources as dust-reddened high-redshift QSOs that might go undiscovered on the basis of the grism data alone.

With these motivations we have undertaken a color analysis of the galaxies and point sources in the entire NICMOS parallel image archive for all fields lying above the Galactic plane ($|b| > 20°$). Our main goal is to use these data to search for two classes of object of strong current interest, specifically, 1.) Extremely Red Objects (EROs; see Thompson et al. 1999, Yan et al. 2000 and references therein), which appear to be very reddened and/or evolved galaxies at $z > 1$, and 2.) dust-reddened and/or very high redshift ($z > 8$) QSOs. By their nature, such objects are difficult to detect in the optical, further motivating a search for them in the NICMOS archive. In the following section we discuss the NICMOS parallel images. In § 3 we discuss our expected results and data simulation. In § 4 we present our analysis and results, and conclude with a discussion in § 5.

## 2. THE NICMOS PARALLEL IMAGES

NICMOS parallel imaging was conducted in Cameras 1 and 2 from 1997 June 2 to 1997 November 8 due to the Camera 3 focus problem that developed shortly after the installation of the instrument (see Thompson et al. 1998). Parallel imaging was then switched to Camera 3 as it came into better focus, due to its larger field of view (51.2″ × 51.2″) and the availability of its grisms. Parallel imaging was continued with Camera 3 until NICMOS exhausted its cryogen supply in 1999 January. Our analysis is of the set of images obtained as part of the "public" parallel program undertaken by the Space Telescope Science Institute, and are available without restriction in the HST archive. The associated HST proposal numbers are 7676, 7726, 7729, 7701, 7811, 7812, 7907, 7922, 8063 and 8082. We use the optimally-reduced Camera 1 and Camera 2 images prepared for the Paper I study, and the "pipeline"-reduced versions of the Camera 3 images available in the HST archive, as the photometric quality of the latter was found to differ from the former by at most a few percent (Paper I). The Camera 3 parallel images were found to show varying amounts of residual amplifier glow at their corners, and for this reason they were trimmed from their original size of 51.2″ × 51.2″ to approximately 39″ × 39″ in order to provide a better estimate of the image background level.

The Camera 2 parallel images were taken in the F110W, F160W and F222M filters, which are analogs of the Johnson-Cousins *JHK* filters (the specific central wavelengths and filter widths in microns are 1.102, 0.592, 1.593, 0.403 and 2.216, 0.143, respectively). However, the higher thermal background in the F222M filter images along with the smaller filter bandpass reduces their sensitivity limit by roughly an order of magnitude relative to the F110W and F160W filters, making them of limited use for a photometric analysis (but see Paper I). We therefore restrict our analysis to the F110W and F160W images. The Camera 1 images were taken only in the F110W and F160W filters, while the Camera 3 images were taken in these two filters plus the grism filters G096 and G141. Integration times were approximately the same for each filter and vary from 92 seconds to 1280 seconds. Further discussion of the image sensitivity limits and completeness is provided in § 4.

We restrict our analysis to fields above the Galactic plane ($|b| > 20°$) for the following reasons. First, fields in the plane are often crowded and suffer from source confusion, even with the high spatial resolutions of the NICMOS detectors. Fields above the plane suffering from source confusion (such as those covering portions of Galactic globular clusters or Local Group galaxies) were also excluded from the analysis. Second, extinction and reddening corrections low in the plane and towards the Galactic center are highly uncertain, even with the most current reddening maps (Schlegel, Finkbeiner & Davis 1998). Finally, as discussed in the next section, the objects most easily distinguished on the basis of their F110W - F160W color are high-redshift galaxies and QSOs, and so require a search at high Galactic latitudes. Figure 1 shows an Aitoff map of the included fields. In Figure 2 we show a representative Camera 3 parallel image obtained in the F160W filter.

## 3. EXPECTED RESULTS AND DATA SIMULATION

### 3.1 Point Sources

In order to interpret the F110W - F160W colors of galaxies and point sources in the parallel fields, we measured this color for a variety of objects using a combination of theoretical models and observational data. For stars we used recent Kurucz model atmospheres (Kurucz 1992) over a temperature range of 4,000° to 60,000° K, and for surface gravities corresponding to the main sequence. Our procedure for these and other spectra was to convolve the spectra with the transmission curves (including detector quantum efficiency) of the two filters, and then measure the F110W - F160W color on the Vega magnitude scale. The Kurucz models of low-temperature stars are relatively uncertain in the near infrared, so we also measured NICMOS colors for the T ~ 3,000° M6 dwarf star Gl 406, using an infrared spectrum provided by Richard Joyce of Kitt Peak National Observatory. For brown dwarfs, we measured colors from the set of models calculated by Burrows et al. (1997) that have fluxes high enough to be detected in the parallel images to within ~ 10 pc. Since, as the models of Burrows et al. show, the spectra and luminosities of brown dwarfs depend in a complex way on their masses and age, not all such objects can be detected to within this distance limit in the parallel fields. For the range of brown dwarf models investigated, the F110W - F160W colors were found to fall in the approximate range for stars in spectral classes G through M, and so their identification requires another color. A similar result obtains for the newly discovered spectral class of L stars, based on translating the $J - H$ colors of the objects discovered by Kirkpatrick et al. (1999) to those of the F110W and F160W filters. Certain rare stars such as Mira variables and extreme carbon stars can have redder colors than the coolest stars we have considered (Whitelock et al. 1994, 1995), but the chances of finding such objects in the relatively small total area and high Galactic lattitudes covered by our sample are small.

To estimate the F110W - F160W colors of QSOs, we used the composite spectrum of ~ 1,000 objects discovered in the Large Bright Quasar Survey (LBQS) presented by Francis et al. (1991) as a template. We first extended this template to include the H$\alpha$ line and surrounding continuum by adding on the red portion of a spectrum of a low-redshift QSO that has a broad H$\beta$ profile similar to that of the composite. The LBQS selected objects with $B_J$ magnitudes < 18.7, and which thus at intermediate and high redshift have a strong rest-frame ultraviolet continuum. However, a study of the optical/infrared colors of Parkes quasars by Webster et al. (1995) has suggested that a potentially large fraction of the QSO population has significantly redder rest-frame UV/optical continua than objects discovered in optical surveys such as the LBQS, and thus that optical surveys for UV-excess objects such as the LBQS may have missed this portion of the QSO population. The origin of the redder continuum of the Parkes quasars compared to optically-selected QSOs is controversial: Webster et al. (1995) suggest that it is produced by dust internal to the QSO host galaxy, while Serjeant & Rawlings (1996) and Benn et al. (1998) have argued that it is due to an increase in the rest-frame optical and near-infrared emission of the objects due to continuum beaming and/or host galaxy starlight, effects that should be strongest among quasars (as opposed to radio-quiet QSOs). It is nevertheless both warranted and possible to use a QSO spectral template that includes the effect of internal dust. To do this we created an artificially reddened version of the LBQS composite spectrum, using a simple $\lambda^{-1}$ dust extinction law. Specifically, we applied such an extinction law to the LBQS composite until its observed-frame *B* band flux at $z \sim 1$ was reduced by an amount comparable to the mean difference between LBQS QSOs and Parkes quasars at this redshift as found by Webster et al. (1995). Thus even if dust is not the dominant source of QSO continuum reddening, this template can still provide an effective way to identify red QSOs.

The reddened and unreddened QSO spectral templates are shown in Figure 3, in the object rest frame. We redshifted these templates over the range $z = 1.7$ to $z = 10$ and measured the F110W - F160W color, approximating the Ly$\alpha$ forest by a zero-flux continuum below 1140 Å in the QSO rest frame, which should be appropriate for the very high redshifts at which the forest enters the F110W filter. The results are shown in Figure 4, where we also include the F110W - F160W colors of three QSOs at $z < 0.2$ from the published *JH* magnitudes of Hyland & Allen (1982), after converting these magnitudes to the corresponding F110W and F160W values. Figure 4 also displays the range of F110W - F160W colors found for the star and brown dwarf models. While unreddened QSOs fall within the color range of stars and brown dwarfs (at least at $z < 8$), as do the QSOs at $z < 0.2$, the reddened QSOs almost immediately become distinguishable from the other classes of object, differing by approximately one magnitude from their unreddened counterparts and by approximately half a magnitude from the reddest stars beginning at $z \cong 1.7$. Of course, with the likely continuous distribution of internal reddening values, as well as the intrinsic variance in the unreddened spectra, a wider spread of object colors is expected.

### 3.2 Galaxies

The NICMOS parallel fields contain galaxies covering a wide range of morphologies and brightnesses, extending in redshift from $z \sim 0$ to possibly as high as $z = 2.7$ (Paper I; McCarthy et al. 1999; see also Figure 2). In Paper I it was found that the stellar population synthesis models of Bruzual & Charlot (1993) could, when combined with ground-based optical photometry, be used to fit the spectral energy distributions of the majority of galaxies in the parallel fields and estimate their photometric redshifts and ages. We thus used the most recent (1995) versions of the Bruzual & Charlot models to simulate the F110W - F160W colors of the galaxies in the parallel fields as a function of redshift. Specifically, we used the models for the case of an instantaneous burst of star formation with input parameters of solar metalicity and a Salpeter initial mass function with a range of 0.1 $M_\odot$ to 125 $M_\odot$, evolved to ages of 0.1 Gyr, 1 Gyr, and 10 Gyr. These models were then redshifted from $z = 0$ to $z = 4$ and the F110W - F160W colors measured. Objects $\sim 10$ Gyr old at $z > 1$ of course violate plausible cosmologies, but their colors above such redshifts were calculated for the sake of completeness.

We also used the spectral energy distribution (SED) of the best-studied ERO, HR 10, as a template for the identification of these objects. Specifically, taking a simple linear interpolation of the *BIJHK* fluxes of this object measured by Graham & Dey (1996) and using $z = 1.44$ (Dey et al. 1999), we measured the F110W - F160W colors of the template over the redshift range $z = 1 - 4$. The very red color ($I - K \cong 6$) of HR 10 appears to be the result of strong internal dust extinction rather than an old stellar population (Graham & Dey 1996; Cimatti et al. 1997, 1998; Dey et al. 1999), and the object appears to be an analog of low-redshift ultraluminous infrared galaxies (ULIRGs) such as Arp 220. By contrast, the Bruzual & Charlot models used do not include the effects of internal dust, and thus represent the opposite extreme of galaxies whose colors are entirely determined by the age of their stellar population. This is unrealistic, given that nearly all well-studied high-redshift galaxies show evidence for dust extinction at some level (e.g. Corbin et al. 1998; Sawicki & Yee 1998; Armus et al. 1998), but as found in Paper I the Bruzual & Charlot models can provide a useful guide for detecting evolved galaxies in which dust extinction is not strong. Spectroscopic confirmation of evolved galaxies at $z \sim 1$ without strong dust extinction has already been obtained (Stockton, Kellogg & Ridgway 1995; Dunlop et al. 1996; Cohen et al. 1999), while photometric evidence of such galaxies at $z > 1.5$ has been found in the Camera 2 parallel fields (Paper I) and in the Hubble Deep Fields (Driver et al. 1998; Benítez et al. 1999).

The tracks of F110W - F160W color versus redshift for these templates are shown in Figure 5, where the range of this color for stars and brown dwarfs has also been marked. While the color tracks show considerable degeneracy between redshift and age, galaxies more than a few Gyr old should become distinct from stars at $z > 1.5$, while EROs like HR 10 become distinguishable from stars and other galaxies at $z > 1.2$. Furthermore, the much redder color of the HR 10 template compared to the dust-free Bruzual & Charlot models out to $z \cong 2.5$ provides at least a preliminary way to distinguish dust-reddened but star-forming galaxies from galaxies with evolved stellar populations.

## 4. PHOTOMETRY

We performed batch photometry on the F110W and F160W parallel images using the SExtractor program of Bertin & Arnouts (1996). Experimentation with subsets of the Camera 2 and Camera 3 sample images showed that a detection threshold set at three times the standard deviation of the image background and no image filtering produced the fewest spurious detections and the detection of all sources obvious to the eye. We additionally required that five contiguous pixels be above this threshold for an object to be considered as a candidate point source or galaxy. This avoids the inclusion of the residual cosmic rays and hot pixels that often persist in reduced NICMOS images (e.g. Figure 2). Comparison of the SExtractor output files for individual fields revealed good but imperfect agreement between the SExtractor object classifications (galaxy versus star) and our visual inspection of the images. In particular, the faintest stars in the images are sometimes classified as galaxies. For the Camera 1 images, SExtractor could not adequately distinguish between stars and galaxies, which is likely due to the more complex point-spread function associated with this high-resolution camera. We therefore excluded the Camera 1 images from the remaining analysis, noting that all galaxies in these fields were already included in the Paper I study, and that due to the small field size, the high Galactic latitude Camera 1 images collectively contain few point sources.

We generated a master catalog of error-free detections from the SExtractor output files from the Camera 2 and Camera 3 images. For galaxy candidates we additionally required that they be unambiguously classified as galaxies (with a SExtractor classification index of < 0.1) in at least one of the filters. For point sources, we required that they be unambiguously classified as such (with a SExtractor classification index of > 0.9) in both filters, in order to avoid the inclusion of spurious features and also to reduce confusion with compact galaxies. We further excluded all objects whose combined error in the F110W and F160W instrumental magnitudes exceeded 50%, in order to limit the F110W - F160W color values to a meaningful range. We used the instrumental magnitudes judged by the SExtractor program as providing the best measurement of the total object flux; for unblended objects, which comprise the majority of those detected, these are the magnitudes measured from the first moment of the object's intensity profile, while for blended objects these are corrected isophotal magnitudes (see Bertin & Arnouts 1996). Transformation of these instrumental magnitudes to a Vega scale was accomplished using the NICMOS on-orbit photometric calibrations (see www.stsci.edu/instruments/nicmos and Colina & Rieke 1997). Corrections for Galactic extinction were made using the reddening maps of Schlegel, Finkbeiner & Davis (1998) and an interpolation of the near infrared reddening law of Rieke & Lebofsky (1985). A small fraction of the fields overlap (Figure 1), either completely or partially, and all overlapping Camera 2 parallel images were shifted and averaged together to increase the image depth (Paper I). Redundant detections in the Camera 3 fields were either discarded on an individual basis or allowed to remain in the object catalog.

In Table 1 we present a summary of the SExtractor output for the total sample of images, including the total area covered and the mean limiting magnitudes and their standard deviations in each filter. The exclusion of all objects with uncertainties in the F110W - F160W color exceeding 50% reduced the effective limiting magnitudes in both filters to be approximately 22. We also note that in the case of the galaxies we have applied a correction to the F110W - F160W color using the $3\sigma$ image background sensitivity limits measured by SExtractor for each field. Specifically, despite the higher sensitivity of the F110W images on the Vega scale (Table 1), these images are in fact slightly less sensitive in terms of flux level than those obtained in F160W, due mainly to shorter integration times. This can lead to erroneous colors for galaxies, as the faint outer regions may go undetected in the F110W filter. The mean value of the associated correction factor to F110W - F160W was approximately -0.3 magnitudes. We also note that the final SExtractor magnitudes for galaxies in the Camera 2 images were found to be in good agreement with those measured for the Paper I study via interactive means, which gives us confidence in the accuracy of the former.

## 5. RESULTS

### 5.1 Galaxies

Our final color-magnitude diagram for the 358 galaxy candidates in the combined Camera 2 and Camera 3 fields is shown in Figure 6. The uncertainties in the color and magnitude values as measured by SExtractor are relatively large, as shown by the mean error bar, and increase with increasing magnitude. Comparison of this plot with Figure 5 shows that we have detected both candidate reddened and evolved objects at $z > 1$ as well as very young (~ 0.1 Gyr) objects to z ~ 1. Specifically, the reddest objects have a F110W - F160W color comparable to or exceeding that of HR 10 at its redshift of 1.44. It seems reasonable to assume that, like HR 10, many if not all of these objects would be classified as EROs under the $R - K' \geq 6$ criterion generally used in ground-based surveys (e.g. Thompson et al. 1999). A difference in mean magnitude between these reddest and bluest galaxies is also evident, which is a consequence of the rough equivalence of the sensitivity levels of the F110W and F160W filters, i.e., to detect the reddest galaxies at $F160W \cong 21.5$ would require that the F110W images go ~ 2 magnitudes deeper than the F160W images. We searched for but did not find objects that dropped out completely in the F110W image but were clearly present in the F160W image.

Considering first the reddest galaxies (arbitrarily limited from Figure 6 to those having F110W - F160W ≥ 1.5), from the 14 objects with such color we find a surface density comparable to that estimated for EROs by Thompson et al. (1999) based on the color selection criterion of $R - K' \geq 6$. Specifically, to a $K'$ magnitude limit of 19.0, Thompson et al. estimate an ERO surface density of approximately 0.04 arcmin$^{-2}$ (140 deg$^{-2}$) whereas to our somewhat deeper limit (F160W $\cong$ 21, modulo the variance in field depth and difference in source detection limits and image resolutions) we find 0.15 arcmin$^{-2}$. This surface density also agrees with that estimated by Yan et al. (2000) for EROs selected under the criterion $R -$ F160W > 5 down to a comparable limiting magnitude in a subset of the NICMOS Camera 3 parallel fields for which they obtained ground-based optical observations. The correspondence between the individual EROs in our sample and those identified by Yan et al. (2000) is not good: there is only one object in common in our set of objects with F110W - F160W ≥ 1.5 and their objects having $R$ - F160W > 5. This is probably the result the different selection criteria used by Yan et al. (2000), and their ERO sample also extends to a higher magnitude limit than ours (F160W > 21.2). Specifically, their sample of EROs is based on visual identification of objects from combined optical and NICMOS images, whereas we have relied on an initial conservative SExtractor classification and include corrections for interstellar reddening and to the F110W - F160W color. Although these results indicate that identifying EROs is problematic, the important finding is the agreement between the estimates of the surface densities of these objects from the different studies. These surface densities are also high (> 100 deg$^{-2}$ to $H \cong 21$) and comparable to those of QSOs (e.g. Koo & Kron 1988; Hall et al. 1996), establishing EROs as a fundamental part of the galaxy population at their probable redshifts ($1 < z < 2$).

In Table 2 we provide a summary of information on all 14 of our ERO candidates, including the object J2000 coordinates and half-light radii. Two of these objects, NPF 003804.90-021602.8 and NPF 154927.03+212111.3, were also included in the Paper I study, and have photometric redshifts of 0.9 (±0.1) and 1.1 (±0.1) respectively, based on the fitting of dust-free Bruzual & Charlot population synthesis models. In Figure 7 we show the F160W images of all 14 objects; considerably less detail is revealed in the F110W images. While the marginal resolution and low signal-to-noise levels of the objects makes interpretation difficult, a variety of morphologies is indicated. In particular, the object NPF 133845.84+701151.4 is clearly a spiral galaxy seen almost perfectly edge-on. For this reason we suspect that its red color is not intrinsic, but is more likely the result of obscuration of the galaxy bulge by dust in the disk. Among the remaining galaxies, there is a suggestion of dust and dynamical disturbance in some, while others appear to be more symmetric and elliptical in appearance. Examples of the former type of galaxies include the reddest object in the sample, NPF 164320.43+171117.4, as well as NPF 012530.99-001129.3 and NPF 115606.07+551440.8, while examples of the latter type include NPF 121839.60+471332.5 and NPF 234358.12-152503.0. With the exception of the edge-on spiral, the morphologies of all the objects having F110W - F160W > 2 suggest the presence of dust, which would be consistent with the colors of the HR 10 template (Figure 5). The positions of 10 of the 14 objects are covered in the FIRST radio survey at 20 cm (Becker, White & Helfand 1995), but a search of the FIRST catalog revealed that none are detected above a ~ 1 mJy flux level. Deeper and higher-resolution near-infrared images of these objects are clearly important for determining their morphologies, and can be obtained if the NICMOS instrument is brought back into normal operation as is currently planned.

All of the blue galaxies in the sample (those with F110W - F160W < -0.2) were inspected and were uniformly found to be very compact and marginally resolved sources. Four examples are shown in Figure 8. Their colors, faintness and compactness suggest that they are blue compact dwarf galaxies at redshifts of at most $z \sim 1$, and we estimate the surface density of these objects to the effective limiting magnitude of F160W $\cong$ 22 to be roughly 300 deg$^{-2}$. We thus find no cases of spiral or elliptical galaxies whose emission is dominated by recent (< 0.1 Gyr) star formation, a result also found in Paper I. No galaxies this blue were found in the subset of the Camera 3 parallel images analyzed by Yan et al. (2000), but since that subset covers only ~ 20% of the area of our sample, this may be the result of cosmic variance, noting that we only find 13 such sources in our larger sample. This may also be the result of different selection criteria used with the SExtractor program in the respective analyses of the images.

## 5.2 Point Sources

Our final color-magnitude diagram for the point sources detected is shown in Figure 9. As with the galaxies, the uncertainties for the faintest objects are large. Comparison with Figure 4 shows that there are no objects with F110W - F160W colors matching those measured for the reddened LBQS spectral template, nor any candidate QSOs at $z > 8$. All the sources detected have colors consistent with being either stars or unreddened QSOs at $z < 7$. Furthermore, there is no indication of a wide range of color among the reddest objects in the sample, as might be expected for QSOs with a large range of internal reddening.

Using this result to establish a limit on the surface density of reddened QSOs is problematic because of the variance in the image depths (Table 1). Also, these detections suffer from the same problem as the galaxy candidates in that the limiting magnitude is color-dependent (Fig. 6). As noted in § 4, SExtractor also occasionally classifies faint stars as galaxies (although none of the very red or blue objects identified as galaxies were found to have a stellar appearance). We assume the effective limiting magnitude for the reddest objects to be approximately the same as that found for the galaxy candidates, F160W $\cong$ 21. We conservatively estimate from the total coverage of the fields that the surface density of QSOs as red as our template is $< 50$ deg$^{-2}$ to this mean limiting magnitude. The apparent optical luminosity function of dust-reddened QSOs will of course differ from UV-excess QSOs, but this should not strongly affect our results, given that even out to $z \sim 2$ the F110W and F160W filters cover the rest-frame optical emission of the objects, which is not as strongly reduced by dust as the rest-frame near-UV (Fig. 3). Importantly, the flux level corresponding to F160W = 22 is approximately ten times lower than that corresponding to the *B* magnitude limits of $\sim 22 - 23$ reached in large optical color surveys for UV-excess QSOs (e.g. Koo & Kron 1988; Hall et al. 1996), which find surface densities of such objects to be $\sim 100 - 200$ deg$^{-2}$. Therefore, while a firm limit on the surface density of a reddened QSO population is not possible, we can conclude that they show no evidence of outnumbering the population of objects selected on the basis of UV excess. Regarding unreddened QSOs at $z > 8$, while any such objects will of course be faint, our magnitude limit still allows us to detect the most intrinsically luminous objects known out to such redshifts, such as the bright $z \cong 5$ sources recently discovered in the Sloan Digital Sky Survey (Fan et al. 1999).

## 6. DISCUSSION

### 6.1 Galaxies

It now appears that EROs, as defined by single infrared and optical-infrared color limits such as $R - K' > 6$, include both ULIRG-analogs such as HR 10 and evolved early-type galaxies at similar redshifts ($1 < z < 2$). For example, Cohen et al. (1999) spectroscopically confirm four objects with $R - K > 5$ to be evolved galaxies at $z \sim 1$, while Smail et al. (1999) find that several submillimeter sources have ERO counterparts, indicating them to be objects similar to HR 10. Our results are consistent with the interpretation of EROs as a mixed population, insofar as 1.) The HR 10 and Bruzual & Charlot spectral templates for ages $\geq 1$ Gyr have the same F110W - F160W colors until $z \sim 2$ (Figure 5), and 2.) The objects in our sample having F110W - F160W > 1.5 show a variety of morphologies, including symmetric sources that may be normal elliptical galaxies as well as those with more irregular morphologies possibly produced by internal dust and interaction.

Multiwavelength observations, such as those made of HR 10, are clearly necessary to break the degeneracy between red colors produced by dust obscuration versus those produced by a very evolved stellar population, and to determine the proportion of the two types of object in the ERO population. In particular, as shown in the case of HR 10, strong submillimeter emission will signify vigorous dust-obscured star formation, while the lack of such emission will indicate an old stellar population. Such observations, in conjunction with further spectroscopy of EROs using very large ground based telescopes (Cohen et al. 1999; Dey et al. 1999) will also allow the determination of the space densities of these subtypes. The determination of the space density of ULIRG-analogs at high redshift is very important for current efforts to measure the evolution of the cosmic star formation rate (e.g. Madau, Pozzetti &

Dickinson 1998), because such efforts have primarily been based on optically-selected galaxy samples which are biased towards objects with strong rest-frame UV emission. Similarly, the spectroscopic confirmation and age-dating of evolved elliptical galaxies at $z > 1$ has important implications not only for models of the formation of these objects, but for the values of the fundamental cosmological parameters (see Stockton et al. 1995; Dunlop et al. 1996).

The fraction of galaxies in the sample that appear to be blue compact dwarfs (BCDs) is roughly 10%. A more precise estimate is difficult, given that some of the galaxy candidates may be misidentified stars, and also because of the large uncertainties in the object colors at the magnitudes of the BCD candidates (Fig. 6). Nevertheless, this fraction of BCDs is comparable to that estimated from combined optical and infrared data for galaxies in the Hubble Medium Deep Survey field (Glazebrook et al. 1998). Redshifts for these galaxies will be required to estimate their space density, and compare such estimates to those obtained from ground-based surveys (e.g. Salzer 1989).

### 6.2 Point Sources

The relatively low inferred surface density of objects with colors matching those of the reddened QSO template runs contrary to the suggestion of Webster et al. (1995) that such sources could comprise up to 80% of the total QSO population. In addition to our result, studies of X-ray selected QSOs reveal very few sources in which internal reddening appears to be strong (Boyle & di Matteo 1995; Kim & Elvis 1999). Spectroscopically confirmed QSO candidates at $z < 0.7$ in the infrared 2MASS all-sky survey also have a mean $J - H$ color of approximately 1.0 (B. Nelson, private communication), which is also not indicative of a large amount of dust extinction (see Fig. 4; the $J - H$ color is comparable to F110W - F160W). It therefore appears that the red colors of the Parkes quasars studied by Webster et al. (1995) are not representative of the majority of the QSO population. If, as argued by Serjeant & Rawlings (1996) and Benn et al. (1998), red continua are characteristic of quasars and not radio-quiet QSOs, then the lack of detection of any such objects in the present sample is consistent with the fact that quasars comprise only ~

10% of the total QSO population, with a correspondingly lower surface density (~ 10 deg$^{-2}$) than radio-quiet QSOs (e.g. Kellermann et al. 1989). The apparent lack of dusty QSOs also argues against a possible evolutionary link between high-redshift ULIRGs and the QSO phenomenon (see Surace et al. 2000 and references therein). More precisely, the lack of evidence for dusty QSOs at least suggests that any transition stage between ULIRGs and typical UV-selected QSOs is short (< 1 Gyr), as for a longer transition time one might expect to detect objects with properties intermediate between the two classes.

Given its ubiquity within high-redshift galaxies, the effect of dust on QSO emission should not be entirely discounted, however. In particular, since the apparent luminosity function of any reddened QSOs will differ from that of non-reddened QSOs, a definitive survey for the former using the color criteria established by Webster et al. (1995) must go deep enough to compensate for the amount of extinction produced in the rest-frame UV (Fig. 3), which past surveys were not designed to do. For the present we can conclude that the effect of internal dust on QSO emission does not appear to be strong. Finally, the lack of evidence for any QSOs at $z > 8$ is consistent with the evidence that QSO space density declines sharply at $z > 5$ (Shaver et al. 1996), and argues against scenarios in which QSOs form at very high ($z > 10$) redshifts (see Haiman, Madau & Loeb 1999 and references therein). Specifically, if the surface density of luminous QSOs at such redshifts were comparable to that at the QSO peak epoch between $z \cong$ 2-3 (~ 100-200 deg$^{-2}$), we could have detected at least one such object in this sample.

We thank the referee, Pat McCarthy, for a helpful review of this paper. We also thank Adam Burrows for providing us with his brown dwarf model spectra, Dick Joyce for copies of his M star spectra, Paul Francis and Brant Nelson for discussions of reddening in QSOs, and Pat Osmer and John Salzer for helpful discussions of QSO and blue compact dwarf galaxy surveys, respectively. Betty Stobie, Irene Barg, Howard Bushouse and Ivo Busko provided invaluable assistance with the NICMOS parallel archive and analysis software, and we also thank Heidi Olson for her help in the processing of the images. This work was supported by NASA grant NAG 5-3042 to The University of Arizona.

TABLE 1

SUMMARY OF IMAGE SAMPLE STATISTICS AND SEXTRACTOR OUTPUT RESULTS

___________________________________________________________________________

| | |
|---|---|
| Total Number of Camera 2 Fields (minus overlapping fields): | 161 |
| Total Number of Camera 3 Fields (minus overlapping fields): | 180 |
| Total Area Covered by Combined Camera 2 and Camera 3 Fields: | 92 arcmin$^2$ |
| Mean 3$\sigma$ Point Source Detection Limit in F110W Filter (Vega Magnitudes): | 23.67 ($\sigma$ = 0.47) |
| Mean 3$\sigma$ Point Source Detection Limit in F160W Filter (Vega Magnitudes): | 23.43 ($\sigma$ = 0.52) |
| Number of Objects Classified as Point Sources: | 631 |
| Number of Objects Classified as Galaxies: | 358 |

___________________________________________________________________________

TABLE 2

EXTREMELY RED GALAXIES (F110W - F160W ≥ 1.5)

| Object | F160W Magnitude | F110W - F160W | $r_{1/2}''$ | Camera |
|---|---|---|---|---|
| 003804.90-021602.8 [a] | 20.0 | 2.0 | 0.38 | 2 |
| 012530.99-001129.3 | 18.7 | 1.7 | 0.60 | 3 |
| 073827.22+385640.6 | 19.5 | 1.7 | 0.80 | 3 |
| 085144.08+115118.3 | 20.4 | 1.7 | 0.80 | 3 |
| 115606.07+551440.8 | 19.3 | 2.0 | 0.80 | 3 |
| 121839.60+471332.5 | 20.3 | 1.7 | 0.40 | 3 |
| 121840.58+471316.1 | 20.7 | 1.5 | 0.40 | 3 |
| 133845.84+701151.4 | 18.8 | 2.3 | 1.20 | 3 |
| 134335.44+012949.9 | 19.4 | 1.6 | 0.60 | 3 |
| 154927.03+212111.3 [a] | 21.2 | 1.8 | 0.23 | 2 |
| 164320.43+171117.4 | 19.1 | 2.6 | 0.30 | 2 |
| 210745.34-051930.1 | 19.7 | 1.8 | 0.40 | 3 |
| 232501.81+280925.4 | 20.0 | 1.7 | 0.40 | 3 |
| 234358.12-152503.0 | 19.0 | 1.6 | 0.80 | 3 |

[a]Additional information including photometric redshift provided in Paper I.

24FIGURE CAPTIONS

FIG. 1. -- Aitoff map in Galactic coordinates showing the distribution of the NICMOS parallel fields included in the present analysis. Crosses represent the Camera 2 fields, open squares the Camera 3 fields. None of the Camera 2 and Camera 3 fields are spatially coincident, even though this is suggested by the placement of some of the fields on the map.

FIG. 2. – Sample Camera 3 parallel image, taken in the F160W filter. The coordinates are J2000 for the field center. The image has been trimmed from $51.2'' \times 51.2''$ to approximately $39'' \times 39''$ to avoid regions at the corners affected by detector amplifier glow.

FIG. 3. -- Composite QSO spectra (shown in the rest frame) used as templates to simulate the colors of QSOs in the F110W and F160W filters. The upper spectrum is the composite formed by Francis et al. (1991) from objects discovered in the Large Bright Quasar Survey, with the red portion of the spectrum of a low-redshift QSO added on to extend the wavelength coverage of the composite to H$\alpha$. The bottom spectrum is the same composite after including the effect of internal dust extinction using a $\lambda^{-1}$ extinction law, with the amount of extinction chosen to yield an approximate $B - K$ color of 6 at $z \sim 1$ (see text).

FIG. 4. -- Comparison of the F110W - F160W color of QSOs with redshift for both the reddened and unreddened versions of the LBQS composite spectrum shown in Figure 3. The comparison also includes the $J - H$ colors of three low-redshift QSOs from Hyland & Allen (1982). The estimated range of this color for stars and brown dwarfs is also shown.

FIG. 5. – Comparison of simulated F110W - F160W color with redshift for galaxies, using Bruzual & Charlot population synthesis models for the case of an instantaneous starburst evolved to ages of 0.1 Gyr, 1 Gyr, and 10 Gyr. The simulated F110W - F160W color of the Extremely Red Object HR 10 is also shown, using a spectral template derived from the broad-band photometry of Graham & Dey (1996) and assuming their redshift value of $z = 1.44$ to be correct. The horizontal lines on the plot indicate the range of F110W - F160W color expected for stars and brown dwarfs.

FIG. 6. – Color-magnitude diagram of all galaxy candidates identified by the SExtractor program. The uncertainties in the values vary with magnitude; the mean errors in both quantities are shown in the upper left corner of the plot.

FIG. 7. – F160W images of the 14 Extremely Red Galaxy candidates listed in Table 2. Images are 3″ square and their orientation is random.

FIG. 8. -- F110W images of four of the galaxy candidates having F110W - F160W < -0.2. Images are 3″ square and their orientation is random.

FIG. 9 – Color-magnitude diagram of all point sources identified by the SExtractor program. The mean error in both quantities is shown in the upper left corner of the plot.